\documentstyle[prd,aps,graphicx]{revtex}

\newcommand{\be}{\begin{equation}}
\newcommand{\ee}{\end{equation}}
\newcommand{\bea}{\begin{eqnarray}}
\newcommand{\eea}{\end{eqnarray}}

\begin{document}
\draft

\title{Dynamics of M-Theory Cosmology}
\author{Andrew P. Billyard}
\address{Department of Physics, Dalhousie University, Halifax, NS, B3H
  3J5, Canada}
\author{Alan A. Coley}
\address{Department of Mathematics and Statistics,
  Dalhousie University, Halifax, NS, B3H 3J5, Canada} 
\author{James E. Lidsey}
\address{Astronomy Unit, School of Mathematical Sciences,  
Queen Mary \& Westfield, Mile End Road, London, E1 4NS, UK}
\author{Ulf S. Nilsson}
\address{Department of Applied Mathematics, University of Waterloo,
  Waterloo, Ontario, N2L 3G1, Canada}

\maketitle{

\begin{abstract}
A complete global analysis of spatially--flat, 
four--dimensional cosmologies 
derived from the type IIA 
string and M--theory effective actions is presented. A non--trivial 
Ramond--Ramond sector is included. The governing equations are
written as a dynamical system. Asymptotically, 
the form fields are dynamically negligible, but play a crucial r\^{o}le in 
determining the possible intermediate behaviour of the
solutions (i.e. the nature of the equilibrium points).  
The only past-attracting solution (source in the system) may be interpreted in the eleven--dimensional 
setting in terms of flat space. This source is unstable to the 
introduction of spatial curvature.
\end{abstract}


}

\section{Introduction}

There are five anomaly--free, perturbative superstring theories \cite{gsw}. 
It is now widely believed that these theories represent special
points in the moduli space of a more 
fundamental, non--perturbative 
theory known as M--theory \cite{Witten95}. (For a 
review see, e.g., Ref. \cite{Sen98}). Moreover, another 
point of this moduli space 
corresponds to eleven--dimensional supergravity. This represents 
the low--energy limit of M--theory \cite{Witten95,Townsend95}.  

The 
original formulation of M--theory was given in terms of the strong coupling 
limit of the type IIA superstring. In this limit, an extra compact  
dimension becomes apparent, with a radius, $R$, related to 
the string coupling, $g_s$, by $R \propto g_s^{2/3}$ \cite{Witten95}. 
The compactification of M--theory on a circle, $S^1$, then 
leads to the type IIA superstring. In this framework, 
the dilaton field of the ten--dimensional string theory 
is interpreted as a modulus field parametrizing the 
radius of the eleventh dimension. 

This change of viewpoint 
re--establishes the importance of eleven--dimensional 
supergravity in cosmology and has interesting consequences for 
the dynamics of the very early universe. An investigation into 
the different cosmological models that can arise in 
M--theory is therefore important and a number of solutions to the 
effective action have recently been found \cite{hw,others,kko,fvm}. 

The bosonic sector of eleven--dimensional supergravity 
consists of a graviton and an antisymmetric, 
three--form potential \cite{CJ79}. 
The purpose of the present paper is to employ the 
theory of dynamical systems to determine the qualitative behaviour of 
a wide class of four--dimensional cosmologies derived from this 
supergravity theory. 
We compactify the theory to four dimensions under the 
assumption that the geometry of the universe is given by
the product $M_4 \times Y^6 \times S^1$, where $M_4$ is 
the four--dimensional spacetime, $Y^6$ represents a six--dimensional,
Ricci--flat internal space and $S^1$ is a circle corresponding to the
eleventh dimension. We assume that the only non--trivial components 
of the field strength of the three--form potential are those 
on the $M_4 \times S^1$ subspace. 

The outline of the paper is as follows. In Section II we 
derive an effective, four--dimensional action 
by employing the duality relationship 
in four dimensions between a $p$--form and a $(4-p)$--form.
The field equations for the class of spatially isotropic 
and homogeneous Friedmann--Robertson--Walker (FRW) universes 
are derived in Section III and expressed as a compact autonomous system
of ordinary differential equations. 
All of the equilibrium points of the system and their stability
are determined in Section IV. 
A complete analysis of the flat cosmological models is presented in
Section V together with a discussion and interpretation of the results.
The robustness of the models is addressed in Section VI wherein a number
of generalizations (i.e. additional degrees of freedom) are included;
in particular curvature effects are considered. We conclude with
a discussion in Section VII.

\section{Four--Dimensional Effective Action}

The bosonic sector of the effective supergravity action for the
low--energy limit of M--theory is given in component form
by\footnote{In this paper, the spacetime metric has signature $(-,+,
\cdots , +)$ and variables in eleven dimensions are represented with
a circumflex accent. Upper case, Latin indices with circumflex accents
take values in the range $\hat{A}=(0, 1, \ldots , 10 )$, upper case,
Latin indices without a circumflex accent vary from $A =(0 , 1, \ldots
, 9 )$, lower case Greek indices span $\mu =(0, 1, 2, 3)$ and lower
case Latin indices represent spatial dimensions. A totally
antisymmetric $p$--form is defined by $A_p = (1/p!) A_{A_1 \ldots A_p}
dx^{A_1} \wedge \cdots \wedge dx^{A_p}$ and the corresponding field
strength is given by $F_{p+1} =dA_p = [1/(p+1)!] F_{A_1 \ldots
A_{p+1}} dx^{A_1} \wedge \cdots \wedge dx^{A_{p+1}}$. The coordinate
of the eleventh dimension is denoted by $\Upsilon$.  The
eleven--dimensional Planck mass is the only dimensional parameter in
this theory \cite{citation} and units are chosen such that $16\pi
\hat{G} =1$.} 
\begin{eqnarray}
\label{Maction}
\hat{S}&=&\int d^{11}x \sqrt{|\hat{g}|} 
\left[ \hat{R} -\frac{1}{48}\hat{F}_{\hat{A}_1\hat{A}_2 
\hat{A}_3\hat{A}_4}\hat{F}^{\hat{A}_1\hat{A}_2
\hat{A}_3\hat{A}_4} \right. \nonumber \\
&&\left. 
- \frac{1}{12^4} \frac{\hat{\epsilon}^{\hat{A}_1\hat{A}_2\hat{A}_3 
\hat{B}_1\hat{B}_2\hat{B}_3\hat{B}_4 
\hat{C}_1\hat{C}_2\hat{C}_3\hat{C}_4}}{\sqrt{|\hat{g}|}}
\hat{A}_{\hat{A}_1\hat{A}_2\hat{A}_3}
\hat{F}_{\hat{B}_1\hat{B}_2\hat{B}_3\hat{B}_4}
\hat{F}_{\hat{C}_1\hat{C}_2\hat{C}_3\hat{C}_4} \right] ,
\end{eqnarray}
where $\hat{R}$ is the Ricci curvature scalar of the eleven--dimensional 
manifold with metric $\hat{g}_{\mu\nu}$, $\hat{g} \equiv {\rm det}
\hat{g}_{\mu\nu}$
and $\hat{F}_{\hat{A}_1\hat{A}_2\hat{A}_3\hat{A}_4} 
\equiv 4\partial_{[\hat{A}_1}\hat{A}_{\hat{A}_2
\hat{A}_3\hat{A}_4 ]}$ is the four--form field strength of the 
antisymmetric three--form potential $\hat{A}_{\hat{A}_1\hat{A}_2\hat{A}_3}$.
The topological Chern--Simons term arises as a necessary 
consequence of supersymmetry \cite{CJ79}. 

In deriving a four--dimensional effective action from Eq. 
(\ref{Maction}), we first consider 
the Kaluza--Klein 
compactification on a circle, $S^1$. This   
results in the effective action for the massless type IIA superstring
\cite{Witten95,effectivecircle}.
The three--form potential $\hat{A}_{\hat{A}\hat{B}\hat{C}}$ 
reduces to a three--form 
potential $A_{ABC}$ and a two-form potential, 
$B_{AB} \equiv A_{AB \Upsilon}$. 
If we ignore the one--form potential that arises 
from the dimensional reduction of the metric, the 
ten--dimensional action is given by \cite{effectivecircle}
\begin{eqnarray}
\label{IIAaction}
S=\int d^{10}x \sqrt{|g_s|} \left[ e^{-\Phi_{10}} \left( R_s +\left( 
\nabla \Phi_{10} \right)^2 -\frac{1}{12} H_{ABC}
H^{ABC} \right) \right. \nonumber \\
\left. -\frac{1}{48}F_{ABCD}F^{ABCD}
-\frac{1}{384} \frac{\epsilon^{A_1A_2 B_1B_2B_3B_4 C_1
C_2C_3C_4}}{\sqrt{|\hat{g}|}} B_{A_1A_2}
F_{B_1B_2B_3B_4}F_{C_1C_2C_3C_4} 
\right] ,
\end{eqnarray}
where $H_{ABC} \equiv 3\partial_{[ A} B_{BC ]}$ 
and $F_{ABCD} =4 \partial_{[ A}A_{BCD ]}$
are the field strengths of the potentials $B_{BC}$ and 
$A_{BCD}$, respectively,  
the ten--dimensional dilaton field, $\Phi_{10}$, is related to the radius 
of the eleventh dimension, $e^{\gamma}$ \cite{Witten95}: 
\begin{equation}
\label{gamma}
\gamma = \frac{1}{3} \Phi_{10}
\end{equation}
and we have performed a conformal transformation 
to the string frame:
\begin{equation}
\label{stringmetric}
g^{(s)}_{AB} =\Omega^2 g_{AB},  \qquad \Omega^2 \equiv 
e^{\gamma} . 
\end{equation}
The first line in Eq. (\ref{IIAaction}) contains the massless excitations  
arising in the Neveu--Schwarz/Neveu--Schwarz (NS--NS) sector of the
type IIA superstring and the second 
line is the Ramond--Ramond (RR) sector of this theory \cite{gsw}. In general, 
the NS--NS fields couple directly to the dilaton field in the string frame, 
but the RR fields do not. 

We now consider the compactification 
of theory (\ref{IIAaction}) to four dimensions. 
The simplest compactification that can be considered 
is on an isotropic six--torus, where the only dynamical 
degree of freedom is the modulus field  
parametrizing the volume of the internal 
space. We therefore 
assume that the string--frame metric (\ref{stringmetric}) 
has the form 
\begin{equation}
ds^2_s =g^{(s)}_{\mu\nu} dx^{\mu} dx^{\nu} +e^{2\beta} 
\delta_{ij} dx^idx^j  ,
\end{equation}
where $\delta_{ij}$ $(i,j =1, \ldots , 6)$  is the 
six--dimensional Kronecker delta and $\beta$ represents the 
modulus field. 

Moreover, we compactify the form--fields in Eq. (\ref{IIAaction}) 
by assuming that the only non--trivial components that remain 
after the compactification are those 
associated with the external spacetime $M_4$. This implies, in particular,
that the Chern--Simons term is unimportant, since 
it is proportional to $F \wedge F$.
The effective four--dimensional action is then given in the string frame by 
\begin{equation} 
\label{4daction}
S=\int d^4x \sqrt{|g_4|} \left[ e^{-\Phi} \left( R+\left( 
\nabla \Phi \right)^2 -6 \left( \nabla \beta \right)^2 -\frac{1}{12} 
H_{\mu\nu\lambda}H^{\mu\nu\lambda} \right) -\frac{1}{48} e^{6\beta} 
F_{\mu\nu\lambda\kappa}F^{\mu\nu\lambda\kappa} \right]  ,
\end{equation}
where
\begin{equation}
\Phi \equiv \Phi_{10} -6 \beta 
\end{equation}
is the four--dimensional dilaton field. 

The field equations and Bianchi 
identities for the form fields are 
\begin{eqnarray}
\label{Hfield}
\nabla_{\mu} \left( e^{-\Phi} H^{\mu\nu\lambda} \right) =0 \\
\label{Hbianchi}
\partial_{[\mu } H_{\nu\lambda\kappa ]} =0
\end{eqnarray}
and 
\begin{eqnarray}
\label{Ffield}
\nabla_{\mu} \left( e^{6\beta} F^{\mu\nu\lambda\kappa} 
\right) =0 \\
\label{Fbianchi}
\partial_{[ \mu } F_{\nu\lambda\kappa\rho ]} =0  ,
\end{eqnarray}
respectively. In four dimensions, a $p$--form 
is dual to a $(4-p)$--form and  
Eqs. (\ref{Hfield}) and (\ref{Ffield}) are solved by the {\em ans\"{a}tze}
\cite{stw91,Qansatz}
\begin{eqnarray}
\label{Hsolution}
H^{\mu\nu\lambda} \equiv e^{\Phi} \epsilon^{\mu\nu\lambda\kappa}
\nabla_{\kappa} \sigma \\
\label{Fsolution}
F^{\mu\nu\lambda\kappa} = Q e^{-6\beta} \epsilon^{\mu\nu\lambda\kappa} ,
\end{eqnarray}
where $\epsilon^{\mu\nu\lambda\kappa}$ is the covariantly 
constant four--form, $\sigma$ is a scalar variable and $Q$ is an 
arbitrary constant.
Although Eqs. (\ref{Hsolution}) and (\ref{Fsolution}) solve 
the field equations (\ref{Hfield}) and (\ref{Ffield}), 
the Bianchi identities (\ref{Hbianchi}) and (\ref{Fbianchi}) 
must also be satisfied. Eq. (\ref{Fsolution}) is trivially satisfied, 
since we are working in four dimensions and substituting 
Eq. (\ref{Hsolution}) into Eq. (\ref{Hbianchi}) implies that 
\begin{equation}
\label{sigmafield}
\nabla_{\mu} \left( e^{\Phi} \nabla^{\mu} \sigma \right) =0  .
\end{equation}

Eq. (\ref{sigmafield}) may be interpreted as the field 
equation for the pseudo--scalar axion field, $\sigma$ \cite{stw91}. 
Moreover, substituting
Eqs. (\ref{Hsolution}) and (\ref{Fsolution}) into 
the remaining field equations for the graviton, dilaton 
and modulus fields implies that they may be derived from a dual 
effective action
\begin{equation}
\label{solitonaction}
S=\int d^4x \sqrt{|g_4|} \left[ e^{-\Phi} \left( R+\left( 
\nabla \Phi \right)^2 -6 \left( \nabla \beta \right)^2 -\frac{1}{2} 
e^{2\Phi} \left( \nabla \sigma \right)^2 \right) 
-\frac{1}{2} Q^2 e^{-6\beta} \right]  .
\end{equation}

In the following Section, we derive the cosmological field equations 
from the effective action (\ref{solitonaction}). 

\section{Cosmological Field Equations}

We denote the FRW metric on $M_4$ by the line 
element $ds_s^2 =- n^2 dt^2 
+e^{2\alpha} d\Omega_k^2$, where $e^{\alpha}$ represents the scale 
factor of the universe, $n$ is the lapse function and $d\Omega^2$ 
is the three--metric on the surfaces of isotropy, 
with positive $(k=+1)$, negative $(k=-1)$ or zero $(k=0)$ curvature, 
respectively. Substituting this {\em ansatz} into the effective 
action (\ref{4daction}), integrating over the 
spatial variables and normalizing the comoving volume to 
unity, yields the reduced action
\begin{equation}
\label{1daction}
S=\int dt \left[ n^{-1}e^{-\varphi} \left( 3\dot{\alpha}^2 
-\dot{\varphi}^2 +6k n^2 e^{-2\alpha} +6\dot{\beta}^2 \right) +\frac{1}{2n}
e^{\varphi+ 6\alpha} \dot{\sigma}^2 -\frac{n}{2} Q^2 
e^{-6\beta +3\alpha} \right] ,
\end{equation}
where a dot denotes differentiation with respect to $t$ and  
\begin{equation}
\label{shifted}
\varphi \equiv \Phi -3\alpha
\end{equation}
defines the `shifted' dilaton field \cite{tsy}.

The corresponding field equations are
\begin{mathletters}
  \label{eq:eveq1}
  \begin{eqnarray}
    \ddot{\alpha} &=& \dot{\alpha}\dot{\varphi} + \frac12\rho - 2k{\rm
    e}^{-2\alpha} - \frac14 Q^2{\rm e}^{-6\beta+\varphi+3\alpha} 
      \label{firsta}\ , \\
  \ddot{\varphi} &=& \frac12\left(3\dot{\alpha}^2 + \dot{\varphi}^2 +
    6k{\rm e}^{-2\alpha} + 6\dot{\beta}^2 - 
    \frac12\rho\right)\ , \\
  \dot{\rho} &=& -6\dot{\alpha}\rho\ , \\
  \ddot{\beta} &=& \dot{\beta}\dot{\varphi} + \frac{1}{4}Q^2{\rm
    e}^{-6\beta + \varphi + 3\alpha} \ , \\
  0 &=& 3\dot{\alpha}^2 - \dot{\varphi}^2 + 6\dot{\beta}^2 +
    \frac12\rho - 6k{\rm e}^{-2\alpha} +  \frac{1}{2}Q^2{\rm
    e}^{-6\beta + \varphi + 3\alpha} \label{eq:hamilt}\ ,
\end{eqnarray}
\end{mathletters}
where we specify $n=1$ and 
\begin{equation}
  \rho = {\rm e}^{2\varphi + 6\alpha}\dot{\sigma}^2 
\end{equation}
parametrizes the kinetic energy of the pseudo--scalar 
axion field. 

Kaloper, Kogan and Olive have considered 
the equivalent compactification of the M--theory effective 
action (\ref{Maction}) directly in terms of eleven--dimensional 
variables when $M_4$ corresponds to the spatially flat FRW spacetime
\cite{kko}. 
In this case, the only 
non--trivial components of the four--form field strength 
that can exist entirely on the subspace $M_4 \times S^1$ are 
$F_{0mnp}$ and $F_{\Upsilon mnp}$, 
where $m=(1,2,3)$, etc. The former represent the non--trivial 
components of the RR
four--form field strength in Eq. (\ref{solitonaction}) and the latter 
are equivalent to those of the NS--NS three--form field strength. The 
scale factors of the universe in the string and M--theory interpretations 
are related by Eqs. (\ref{gamma}) and (\ref{stringmetric}). 
These relationships provide the recipe that allows the type IIA 
string cosmologies to be reinterpreted in terms of eleven--dimensional, 
M--theory models. It can be verified by direct 
comparison that for $k=0$, 
Eqs. (\ref{firsta})--(\ref{eq:hamilt}) are formally equivalent to 
the field equations derived in Ref. \cite{kko}. The advantage of 
employing the string-frame variables in this work is that the first 
derivative of the shifted dilaton field (\ref{shifted}) is a dominant 
variable and this greatly simplifies the analysis 
of the global dynamics. 

To proceed, we define a new time variable, $\eta$: 
\begin{equation}
  \frac{d\eta}{dt} \equiv  {\rm e}^{(-6\beta + \varphi +
  3\alpha) /2} \ .
\end{equation}
The system of equations (\ref{eq:eveq1}) then becomes:
\begin{eqnarray}
  \alpha'' = \frac12\alpha'\varphi' +3\alpha'\beta' -
  \frac{9}{2} \left(\alpha'\right)^2 + \left( \varphi' \right)^2 -
  6\left(\beta'\right)^2 +4k{\rm e}^{-(5\alpha +
    \varphi-6\beta)}  -
  \frac{3}{4}Q^2\ , \\  
  \varphi'' = 3\left(\alpha'\right)^2 + 6\left(\beta'\right)^2 +
  \frac14Q^2 - \frac{1}{2}\left( \varphi' \right)^2 +
  3\beta'\varphi' - \frac{3}{2}\alpha'\varphi'\ , \\
  \beta'' = \frac12\beta'\varphi' +
  3\left(\beta'\right)^2 - \frac{3}{2}\alpha'\beta' +
  \frac{1}{4}Q^2 \ , \\
  \frac12\rho{\rm e}^{-(\varphi+3\alpha-6\beta)} = \left( \varphi'
  \right)^2 - 3\left(\alpha'\right)^2 - 
  6\left(\beta'\right)^2 + 6k{\rm e}^{-(5\alpha +
    \varphi-6\beta)} - \frac{1}{2}Q^2\ ,  
\end{eqnarray}
where a prime denotes differentiation with respect to $\eta$ and 
the Hamiltonian constraint (\ref{eq:hamilt}) has been employed
to eliminate 
the axion field, $\rho$. 

Since $Q^2$ is semi--positive definite, 
it follows from Eq. (24) that $\varphi'$ is a dominant
variable in the spatially flat and negatively curved models $(k \le 0)$. 
In addition, it follows from Eqs. (22) and (24) 
that $\varphi'$ is a monotone
function. This is important because it implies the global
result that $\varphi'$ is either {\it monotonically increasing or  
decreasing} throughout the evolution of the models. (The variable $\varphi'$ 
plays an analogous r\^{o}le to that of the expansion parameter 
in the spatially homogeneous 
perfect fluid models of general relativity \cite{WEBOOK}).

We therefore introduce the new dimensionless time variable, $\tau$, 
according to 
\begin{equation}
  \frac{d\tau}{d\eta} = \varphi'\ , 
\end{equation}
where we assume here that  $\varphi' >0$. (The case $\varphi' < 0$ is
discussed in Section V). 
We also define the following set of dimensionless variables:
\begin{equation}
\label{dimensionless}
  x \equiv  \frac{\sqrt{3}\alpha'}{\varphi'} \ , \ y \equiv 
  \frac{\sqrt{6}\beta'}{\varphi'}\ , \ z \equiv 
 \frac{Q^2}{2(\varphi')^2}\ , \
  u \equiv  -\frac{6k{\rm e}^{-(5\alpha + \varphi-6\beta)}}{(\varphi')^2}\ ; 
	\ \Omega \equiv  \frac{{\rm
  e}^{-(\varphi+3\alpha-6\beta)}\rho}{2\left(\varphi'\right)^2}\ .
\end{equation}
This leads to a decoupling of the equation for $\varphi'$, which can be
written as
\begin{equation}
  \frac{d\varphi'}{d\tau} = \left(-\frac12 + \frac12z -
  \frac{\sqrt{3}}{2}x + x^2 + y^2 +
  \frac{\sqrt6}{2}y\right)\varphi' \ .
\end{equation}
The remaining equations can then be written in the following
dimensionless form:
\begin{mathletters}
  \label{eq:dimless}
\begin{eqnarray}
  \frac{dx}{d\tau} &=& \left(1-x^2-y^2-z\right)(x+\sqrt{3}) +
  \frac12z\left(x-\sqrt{3}\right)- \frac{2}{\sqrt{3}}u \
   , \label{dxdtau}\\ 
  \frac{dy}{d\tau} &=& \left(1-x^2-y^2-z\right)y + \frac12z\left(y +
    \sqrt{6}\right) \ ,\label{dydtau} \\
  \frac{dz}{d\tau} &=& \left[z-1-\sqrt{6}y + \sqrt{3}x +
    2\left(1-x^2-y^2-z\right)\right]z\ , \\
  \frac{du}{d\tau} &=& -\frac{1}{3}u\left[2\sqrt{3}x +
    3(z+2x^2+2y^2)\right].
\end{eqnarray}
\end{mathletters}
The variable $\Omega$ is given by
\begin{equation}
  \Omega = 1 - x^2 - y^2 - z - u,
\end{equation}
and satisfies the auxiliary equation
\begin{equation}
  \frac{d\Omega}{d\tau} = -\left[2y^2 + z +
  2x\left(x+\sqrt{3}\right)\right]\Omega \ .
\end{equation}

\section{Structure of state space and local analysis}

In this Section we present all of the equilibrium points that arise in the 
system (\ref{eq:dimless}). We are primarily interested in the 
spatially flat models. However, we also 
consider the stability of these models
to perturbations in the spatial curvature. The local stability
analysis we perform is valid for both positive
and negative spatial curvature, although we explicitly
consider the $k\leq0$ models since in this case the condition 
$\rho\geq0$ implies that all of the dimensionless variables
are bounded. The physical state space is defined by
\begin{equation}
  0 < \left\{ x^2, y^2, z , u\right\} < 1 
\end{equation}
and a global analysis can therefore be undertaken. 

We include the boundary of the state space in our 
analysis because the dynamics in the invariant boundary
submanifolds is useful in determining 
the global properties of the orbits in the
physical phase space.
The boundary of the state space consists of a number of invariant
submanifolds of the system. They are:  (i) models 
where the axion field is trivial ($\Omega=0$), (ii)
the spatially flat models ($u=0$), and (iii) models where $z=0$, 
corresponding to the case where the four--form field strength, 
$F_{\mu\nu\lambda\kappa}$, is dynamically unimportant. 
The system of equations
also admits an invariant submanifold, $K$, that is not part of the boundary
of state space:
\begin{equation}
  \label{eq:nontriv}
  K: x +\sqrt{2}y +\sqrt{3} = 0\ , \ u=0\ .
\end{equation}

The equilibrium points are: 

\vspace{.1in}

{\em Equilibrium set: The Line} $L^+$
$$
x^2 + y^2 = 1 \ , \ z=0 \ , \ u=0 \ \ ; \ \Omega = 0\ .
$$
$$
\lambda_1=0\ , \ \lambda_2 = -2(\sqrt{3}x+1)\ , \ \lambda_3=-1 +
\sqrt{3}x - \sqrt{6} y \ , \ \lambda_4 = -2 \left( \frac{1}{\sqrt{3}}x
+1 \right)   \ , 
$$
where $\lambda_i$ denote the eigenvalues. 
The zero eigenvalue indicates that this is indeed an equilibrium
set, corresponding to a circle of unit radius in the 
$(x,y)$ plane. We refrain from presenting 
the eigenvectors, but note that it is the
eigendirection associated with $\lambda_3$ that points in a direction
outside the submanifold $z=0$, while the eigenvector of
$\lambda_4$ extends into the $u$ direction. The stability of these equilibria
is discussed in the following Section\footnote{For hyperbolic equilibrium 
points the stability
is determined by the signs of the real parts of the associated 
eigenvalues; in the case of 
a source (past attractor) all are positive and in the case of a 
sink (future attractor) all are negative. Otherwise, 
the point is a saddle. If the real part of any of the 
eigenvalues of an isolated equilibrium
point is zero, it is non-hyperbolic and the stability cannot be determined 
directly from the eigenvalues.}.   

\vspace{.1in}

{\em Equilibrium point: The source} $R$
$$
x = -\frac{5\sqrt3}{19}\ , \ y =
-\frac{\sqrt6}{19}\ , \ z = \frac{28}{361}\
, \ u = \frac{252}{361}\ ; \ \Omega=0 \ .
$$
$$
\lambda_1 = \frac{20}{19}\ , \ \lambda_2 =
\frac{14}{19}\ , \ \lambda_{3,4} =\frac{7 \pm
  i\sqrt{119} }{19}\ .
$$

\vspace{.1in}

{\em Equilibrium point: The saddle} $M$
$$
x=-\frac{1}{\sqrt{3}}\ , \ y = z =0 \ , \ u = \frac{2}{3}\ ; \
\Omega=0\ .
$$
$$
\lambda_1 = \lambda_2 = 
\frac{2}{3}\ , \ \lambda_3 = \frac{4}{3}\ , \ \lambda_4 = -\frac{2}{3} 
$$
The saddle point $M$ 
corresponds to the Milne form of flat space. This may be mapped onto 
the future light cone of the origin of Minkowski spacetime and 
in this sense may be interpreted as the string perturbative vacuum 
represented in terms of non--standard coordinates. 

\section{Dynamics of the Spatially Flat Cosmologies}

\subsection{Global Analysis}

In this Section we consider the global dynamics of the spatially flat 
cosmologies ($k=0$, $u=0$). For these models, the state space is
three--dimensional and the orbits can therefore  be 
represented pictorially. 

The only equilibrium points in the spatially flat models  lie on the line 
$L^+$ and the eigenvalues are given in Section IV.  From these eigenvalues, it can be seen that
this line is a sink for $x>-1/\sqrt3$ and $\sqrt2 y>x-1/\sqrt3$.  The
lines $x=-1/\sqrt3$ and $\sqrt2 y=x-1/\sqrt3$ intersect on $L^+$ at
the point $P: (x,y)=(-1/\sqrt{3},-\sqrt{2/3})$,
at which all three eigenvalues are zero. Hence, $P$ is a non-hyperbolic
equilibrium point. All other points on $L^+$ are saddles. 

It can be shown
that the point $P$ is a source in the three-dimensional phase space. It 
follows from Eqs. (\ref{dxdtau}) and (\ref{dydtau}) that
\begin{eqnarray}
\label{pseudomono}
\frac{d}{d\tau}\left(x+\sqrt2y+\sqrt3\right) = \left(x+\sqrt2 y+\sqrt3\right)
	\left(1-x^2-y^2-\frac{1}{2} z\right) 
\end{eqnarray}
for $u=0$. This implies that $x+\sqrt2y+\sqrt3$ is a {\em
monotonically increasing} function in the physical phase space. 
The term $(1-x^2-y^2-\frac{1}{2} z)$ is positive--definite  
in the interior region and can only be zero on the boundary, 
where $x^2+y^2=1$ and $z=0$. The term $x+\sqrt2y+\sqrt3$ is 
positive--definite in the physical state space and can only 
vanish in the extended phase space at the point $P$. Indeed, 
the line $x+\sqrt2 y=-\sqrt3$ is tangent to
the unit circle $x^2+y^2=1$ and $z=0$ and actually touches it at the point $P$.
We may conclude, therefore, that the non-hyperbolic equilibrium point $P$ is 
indeed a source for the three-dimensional
system. We have verified this by analysing the equilibrium point $P$ 
using spherical polar coordinates and by numerical calculations. 

The dynamics on the boundary of the state space is also important 
when interpreting the behaviour of the orbits. 
The boundary consists of the two invariant submanifolds 
$\Omega=0$ and $z=0$.  The $\Omega =0$ (trivial axion field) 
submanifold can be solved
analytically in terms of the variables of the state space and the solution is
given by 
\begin{equation}
  y = -\sqrt{6} +
  \frac{(y_0+\sqrt{6})(x-\sqrt{3})}{x_0-\sqrt{3}}\ ,
\end{equation}
where $(x_0, y_0)$ represents  the initial point of the orbit. Thus, 
orbits follow straight line paths in the $(x,y)$ plane. Moreover, 
since by definition $x < \sqrt{3}$, this variable is a
monotonically {\em decreasing} function on this submanifold and $y$ 
is a {\em monotonically increasing} function. 
The line $L^+$ is a source for $x> -1/\sqrt{3}$ and 
$\sqrt{2}y < x-1\sqrt{3}$ and a sink otherwise. 

The boundary $z=0$ describes models where the four--form field strength 
is dynamically negligible. This submanifold can also be solved exactly and 
the orbits follow the straight line paths:
\begin{equation}
  y = \frac{y_0(x+\sqrt{3})}{x_0+\sqrt{3}}\ ,
\end{equation}
where $(x_0, y_0)$ again represents the initial point of the orbit. In this 
case, the
function $x$ is monotonically {\em increasing} on this
submanifold. The line $L^+$ is a source for $x < -1/\sqrt{3}$ and 
a sink otherwise. 

The time--reversed dynamics of the $\varphi'>0$
models we have considered thus far 
is equivalent to the dynamics of the case where $\varphi'<0$. This follows by 
redefining the time variable, $\tau$:
\begin{equation}
\label{newtime}
  \frac{d\tau}{d\eta} = -\varphi'\ ,
\end{equation}
so that $\eta$ and $\tau$ are both increasing or both decreasing together.
If we define the other state variables as in Eq. (29), 
the variables $x$ and $y$ for $\varphi'<0$ are now the
reflections of the variables $x$ and $y$ for $\varphi'>0$, 
i.e., $x\rightarrow -x$ and $y\rightarrow -y$.  With the new time
variable (\ref{newtime}), 
the evolution equations (\ref{eq:dimless}) will have an `overall' change in sign, i.e.,
$dx/d\tau\rightarrow -dx/d\tau$, etc.  
Thus, the equilibrium points are identical in both cases, 
but the eigenvalues have opposite signs. 
Consequently, the dynamics 
of the $\varphi ' < 0$ models is the time reversal of the  
$\varphi'> 0$ models, where 
contracting models for $\varphi'>0$ are expanding models for
$\varphi'<0$, and vice versa. 

\subsection{Physical Interpretation}

The phase space for the spatially flat models 
is depicted in Figs. 1--4. Figs. 1 and 2 correspond to 
the invariant submanifolds $z=0$ and $\Omega =0$, respectively. 
Figs. 3 and 4 represent views of two typical orbits in the full 
three--dimensional phase space. 

\begin{figure}[htp]
  \centering
   \includegraphics*[width=3in]{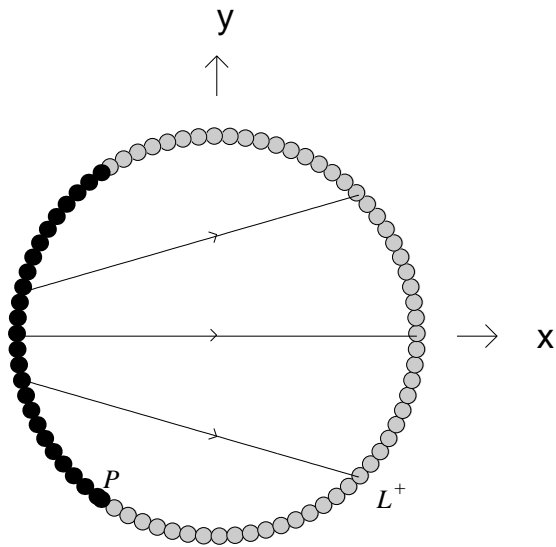}
  \caption{{\em Phase portrait of the invariant submanifold $z=0$, 
corresponding to the case where the
RR four--form field strength is trivial and 
the NS--NS three--form field strength is dynamically 
important. The line $L^+$ represents a
line of equilibrium points. Large black dots denote repellers (sources) 
while grey-filled dots denote attractors (sinks).  
The point $P$ represents a source in
both the two-dimensional and three-dimensional phase spaces. }}
 \end{figure}

\begin{figure}[htp]
  \centering
   \includegraphics*[width=3in]{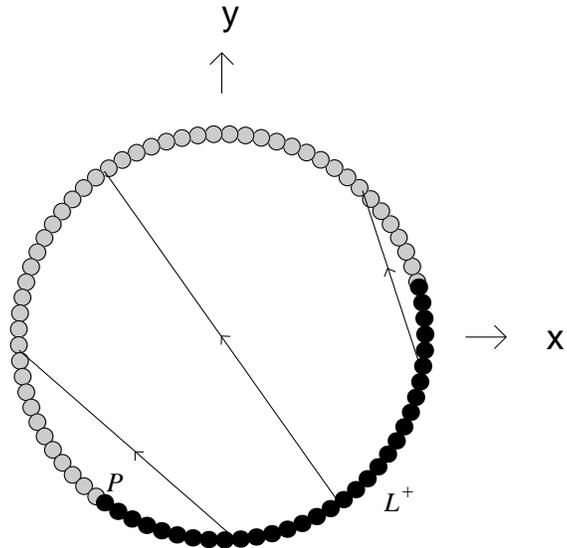}
  \caption{{\em Phase portrait of the invariant submanifold $\Omega =0$, 
corresponding to the case where the NS--NS three--form field 
strength is trivial and the RR four--form field strength is 
dynamically important. See also caption to Figure 1.}}
\end{figure}

\begin{figure}[htp]
  \centering
   \includegraphics*[width=5in]{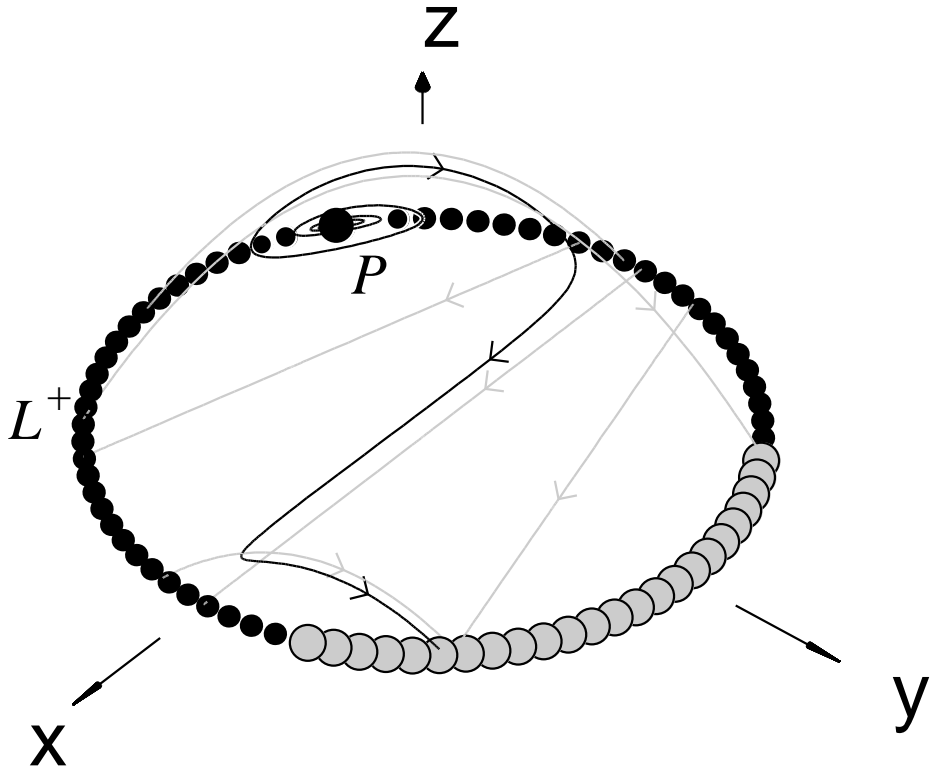}
  \caption{{\em Phase diagram of the spatially flat cosmologies when 
both NS--NS and RR form fields are non--trivial. 
Note that $L^+$ represents a
line of equilibrium points.  The trajectories in Figures
1 and 2 are depicted in grey in this figure along $z=0$
and $z=1-x^2-y^2$, respectively.  Small black dotes represent
saddle points.  See also caption to Figure 1.}}
\end{figure}

\begin{figure}[htp]
  \centering
   \includegraphics*[width=5in]{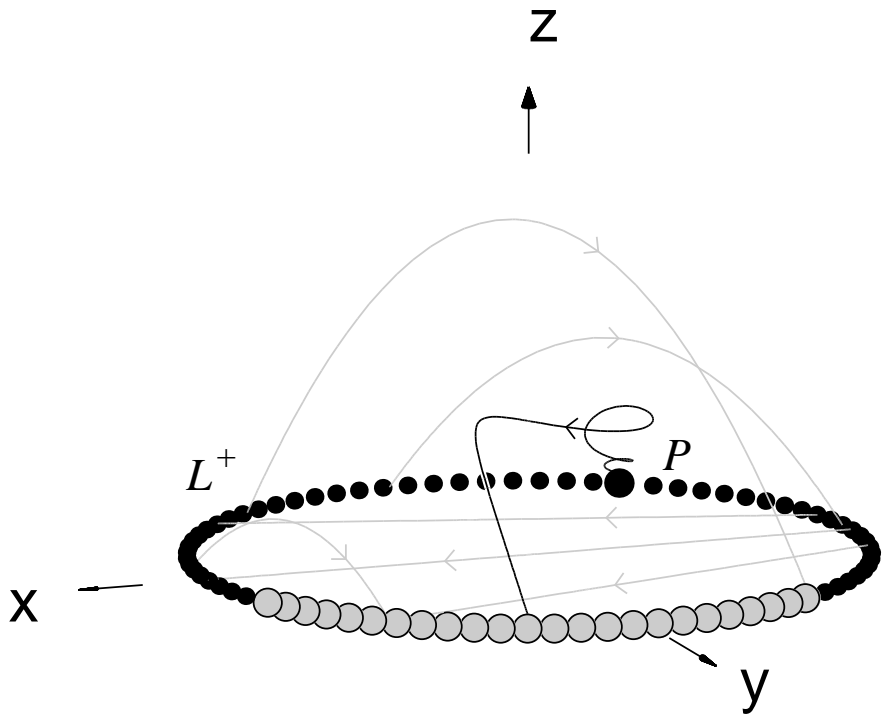}
  \caption{{\em An alternative view of a different trajectory in
the spatially flat phase space.
See also captions to Figures 1 and 3.}}
\end{figure}

The equilibrium set $L^+$ represents solutions where the form--fields 
are trivial and 
only the dilaton and moduli fields are dynamically important. 
These are known as the `dilaton--moduli--vacuum' 
solutions and have an analytical form given by
\begin{eqnarray}
\label{dmv}
e^{\alpha} =e^{\alpha_*} |t|^{\pm h_*} \nonumber \\
e^{\Phi}=e^{\Phi_*} |t|^{\pm 3h_* -1} \nonumber \\
e^{\beta} =e^{\beta_*} |t|^{\pm \epsilon\sqrt{(1-3h_*^2)/6}} ,
\end{eqnarray}
where $\{ \alpha_* , \Phi_* , \beta_* , h_* \}$ are constants, and
$\epsilon=\pm1$. 
Note that the ``$-$'' solutions in Eq. (\ref{dmv}), which are
represented by the line $L^+$, correspond to $t<0$ and, 
in the time-reverse case
($\dot\varphi<0$), the ``$+$'' solutions of Eq. (\ref{dmv}) correspond to
$t>0$. 

Let us first consider the dynamics in the invariant 
submanifold $z=0$, where the NS--NS axion field is non--trivial and the 
RR four--form field strength vanishes (Fig. 1). 
These trajectories represent the `dilaton--moduli--axion' 
solutions discussed in Ref. \cite{coplahwan}. The trajectory along 
$y=0$ corresponds to the solution where the internal dimensions are 
static. In this case, the universe is initially contracting 
$(x < 0)$, but ultimately bounces into an expansionary phase 
($x>0$). The bounce is 
induced by the two--form potential. It follows from Eq. 
(\ref{Hsolution}) that the field strength of this antisymmetric tensor 
field is directly proportional to the volume form of the 
three--space. This implies that the axion field may be 
interpreted as a membrane that is wrapped around the spatial hypersurfaces
\cite{kaloper}. 
This membrane resists the initial collapse of the 
universe and results in a bouncing cosmology. 
Many solutions exhibit such a bounce, but others 
collapse to zero volume. These arise when the initial 
kinetic energy of the modulus field (internal space) is sufficiently high 
that it can always dominate the kinetic energy of the axion field. 

In the other invariant submanifold $(\Omega =0)$, the NS--NS 
two--form potential 
is trivial, and the RR three--form potential is dynamical. The 
cosmological constant term $Q^2$ in the effective action 
(\ref{solitonaction}) may be interpreted as a $0$--form field 
strength. 
In a certain sense, this 
degree of freedom plays a r\^{o}le analogous to that of a domain 
wall\footnote{In general, a solitonic $p$--brane is supported 
by the magnetic charge of a $(D-p-2)$--form field strength in $D$ 
spacetime dimensions.} \cite{cowlupop}. 
However, in contrast to the membrane 
associated with the axion field, this
`domain wall' resists the expansion of the universe. Thus, 
the majority of solutions that are 
initially expanding ultimately recollapse, as shown in Fig. 2. There are some 
solutions where the internal space is initially 
evolving sufficiently rapidly that the modulus field 
dominates the form field and the expansion can proceed indefinitely.  
Solutions that are initially collapsing do not undergo a bounce. 

In both invariant submanifolds, the point $P$ corresponds to an endpoint 
on the line of sources. In Fig. 1, 
the reflection of this point in the line $y=0$ represents the 
opposite end of the line of sources. This point 
corresponds to a dual solution, where the 
radius of the internal space is inverted. 
Thus, the endpoints of the line of 
sources in the invariant submanifold $z=0$ are 
related by a scale factor duality. 

We may now consider the dynamics 
in the full three--dimensional phase space, where both the 
NS--NS two--form potential and RR three--form potential 
are dynamically significant. Although they are asymptotically negligible, 
the interplay between these fields has important consequences. The 
key point is that the RR field causes the universe to 
collapse, but the NS--NS field has the opposite effect. 
These two fields therefore 
compete against one another, as can be seen in Figs. 3 and 4. 

The point $P$ is the only source in the 
system when both form fields are present. 
Furthermore, it follows from the definitions (\ref{dimensionless}) 
that it represents the collapsing, isotropic, ten--dimensional 
cosmology, where $\alpha =\beta$. The four--dimensional 
dilaton field, $\Phi$, is trivial in this case. 
As the collapse proceeds, a typical orbit moves upwards in a cyclical fashion
until a critical point is reached, where 
one of the form fields is able to dominate the dynamics. 
The orbit then shadows 
the corresponding trajectory in the invariant submanifold $z=0$ or 
$\Omega =0$. In Fig. 3, the axion field dominates and 
causes the universe to bounce. By this time, however, the kinetic 
energy of the modulus field has become significant and the 
solution ultimately asymptotes to a dilaton--moduli--vacuum solution 
on $L^+$. 

All sinks in this phase space correspond to 
solutions where the internal dimensions are expanding $(y >0)$.  
There is a particular point where the spatial dimensions spanning 
the spacetime $M_4$ become static in the late--time limit. In general, 
however, solutions either collapse to 
zero volume in a finite time or superinflate $(\ddot{\alpha} >0)$ 
towards a curvature singularity. In this sense, they correspond to 
pre--big bang cosmologies, since the comoving 
Hubble radius decreases  \cite{Veneziano91,pbb}. 
However, since the internal space is 
expanding, it is not clear to what extent this behaviour 
represents a realistic, four--dimensional 
inflationary solution. 

As discussed above, the time-reversed dynamics of the above 
class of models is deduced by interchanging 
the sources and sinks and reinterpreting 
expanding solutions in terms of contracting ones, and vice--versa. Thus, 
the late-time attractor for the time--reversed system is 
the expanding, isotropic, ten--dimensional cosmology located at point $P$. 

It is of interest to reinterpret the equilibrium 
points of the phase space in terms of eleven--dimensional solutions. Since 
the eleven--dimensional three--form potential is trivial on $L^+$, these 
points represent `Kasner' solutions to eleven--dimensional, vacuum 
Einstein gravity. Thus,  
the line $L^+$ is analogous to the Kasner ring that arises in the 
vacuum Bianchi I models of four--dimensional general relativity \cite{WEBOOK}. 

For the compactification we have considered, the scale factors 
in the eleven--dimensional frame
are $\{ e^{\tilde{\alpha}} , 
e^{\tilde{\beta}} , e^{\gamma} \}$, where, from Eq. (\ref{stringmetric}), 
$\tilde{\alpha} \equiv \alpha -\gamma /2$ and $\tilde{\beta}\equiv 
\beta -\gamma /2$. The `Kasner' solutions are then given by 
the power laws $\tilde{\alpha} = \alpha_0 /\tilde{t}$, 
$\tilde{\beta} =\beta_0 /\tilde{t}$ and $\gamma = \gamma_0 /\tilde{t}$, 
where $\tilde{t} \equiv \int dt \exp (-\gamma /2)$ and the constants 
of integration satisfy the constraints 
$3\alpha_0 +6\beta_0 +\gamma_0 =1$ and 
$3\alpha_0^2 +6\beta_0^2 +\gamma_0^2 =1$. 

These redefinitions imply that the source $P$ corresponds to 
the `Kasner' solution $(\alpha_0 , \beta_0 , \gamma_0 ) =(0,0,1)$. 
This point represents the Taub form of Minkowski spacetime 
\cite{WEBOOK}. The relevance of this solution 
to the problems associated with the pre--big bang 
curvature singularity have recently been discussed \cite{kko,fvm}, and 
it is interesting from an 
eleven--dimensional point of view that such a simple solution is 
uniquely selected by the dynamics. 
The endpoints of the line of sinks on $L^+$ correspond 
to the `Kasner' solutions $(1/2,0,-1/2)$ and $(0,2/7, -5/7)$, respectively, 
and consequently in both cases a subset of the scale factors are static.  

This concludes our discussion of the phase space for the spatially 
flat cosmologies. In the following Section, we consider the robustness 
of these models to a number of possible generalizations, including the 
effects of spatial curvature. 

\section{Robustness of the Models}

\subsection{Effects of Spatial Curvature}

Although the compactness of the phase space depends on the fact that
$k \le 0$, 
one can assume arbitrary signs for $k$ in order to determine the {\em local}
stability of the equilibrium points in the three-dimensional set $u=0$
with respect to curvature perturbations.
The eigenvalue associated with $u$ for the equilibrium points $L^+$ is
always negative. This means that the sinks on $L^+$
(i.e., points on $L^+$ for $x>-1/\sqrt3$ and $\sqrt2
y>x-1/\sqrt3$) remain sinks in the 
four-dimensional phase-space. 
In addition, this implies that the point  $P$ is now only a saddle;
that is, {\em the stability of $P$ 
is unstable to the introduction of 
both positive and negative spatial curvature}.

Since a portion of $L^+$ 
acts as sinks in the four-dimensional phase space, there exists the
{\em global} result that the corresponding
dilaton--moduli--vacuum solutions (\ref{dmv})
(for $3\dot\alpha>-\dot\varphi$ and
$6\dot\beta>3\dot\alpha-\dot\varphi$) will be attracting solutions for
the spatially curved models. We may deduce 
further global results by restricting 
our attention to the negatively curved models $(k<0)$, in which case 
the four-dimensional phase space is compact. As discussed above, 
the point $P$ is a only a saddle point in this extended phase 
space. Moreover,  it follows 
from the analysis of Section IV that the only attracting equilibrium 
point is the point $R$. (There is an additional saddle
$M$ which will affect the possible intermediate dynamics). 
This source 
corresponds to a negatively curved model with a trivial axion field; 
indeed, it is a
power-law, self-similar collapsing solution with non-negligible 
modulus and RR form--field. 

We have been unable to find a monotone function on the extended 
four-dimensional
phase space, but it is plausible that
all negatively-curved models evolve from the solution
corresponding to the global source $R$ 
towards the dilaton--moduli--vacuum solutions 
(on the attracting portion of  $L^+$).
Clearly the curvature is dynamically important at early times.

In the time-reverse case, the solutions asymptote from the
non-inflationary dilaton--moduli--vacuum solutions in the past and  evolve to
the future towards a curvature dominated model; it is plausible that they evolve
towards a model which is the time-reversal of the one represented by
$R$.  Therefore, curvature can also be dynamically important at late
times.

\subsection{Effects of Generalized Couplings}

We now consider a generalization of the effective action 
(\ref{solitonaction}) given by 
\begin{equation}
  \label{eq:action_2}
  S = \int d^4x \sqrt{-g}\left\{ {\rm e}^{-\Phi}\left[ R +
  \left( \nabla \Phi \right)^2 - 6 \left( \nabla \beta \right)^2 
  -\frac12{\rm e}^{2\Phi} \left( \nabla \sigma \right)^2 - 2\Lambda\right]
  - \frac{1}{2}Q^2{\rm e}^{c\beta}\right\} ,
\end{equation}
where $\Lambda$ represents  
a cosmological constant term and $c$ is an arbitrary
constant. The former term may arise through 
non--perturbative corrections to the string effective action. 
The motivation for considering an 
arbitrary coupling of the modulus field to the four--form field strength
is that the generality of the dynamics discussed in Section 
IV  (in which $c=-6$) can be investigated. Eq. (\ref{eq:action_2}) reduces to the action 
studied in Ref. \cite{BCL} when $c=0$. 

By invoking the same assumptions as in section III, the action 
(\ref{eq:action_2}) reduces to
\begin{equation}
  S = \int dt\left[{\rm e}^{-\varphi}\left(3\dot{\alpha}^2 -
  \dot{\varphi}^2 + 6k{\rm e}^{-2\alpha} + 6\dot{\beta}^2 -
  2\Lambda\right) + \frac12{\rm e}^{\varphi+6\alpha}\dot{\sigma}^2 -
  Q^2{\rm e}^{c\beta + 3\alpha}\right]\ ,
\end{equation}
and the corresponding field equations can again be derived from this action. 
In analogy with Eqs. (23) and (28), we
introduce a new time variable, $\tau$, defined by
\begin{equation}
  \frac{d}{d\tau} = \left(\varphi'\right)^{-1}{\rm e}^{-\frac12(c\beta 
+ \varphi +
  3\alpha)}\frac{d}{dt} \ ,
\end{equation}
and the new reduced variable
\begin{equation}
  v =\frac{2\Lambda{\rm e}^{-(3\alpha + \varphi +  c\beta)}}{(\varphi')^2}. 
\end{equation}
{}From these definitions and the reduced variables defined earlier, 
we obtain a five-dimensional system of ordinary differential
equations for the reduced (dimensionless) variables after 
eliminating the variable $\Omega$ that is 
now defined by $\Omega \equiv 1 - x^2 - y^2 - z - u - v$. 
Since $\rho\geq0$, all of the dimensionless variables
are bounded for the models with
$k\leq0$ and $\Lambda>0$, where the physical state space is defined by
$0 < \left\{ x^2, y^2, z , u, v\right\} < 1$, and a
global analysis is therefore possible in this case.  
Including the boundaries $\Omega=0$, $z=0$, 
$u=0$ and $v=0$ leads to a compact state space.

We can analyse these models and obtain qualitative information about the
dynamics in an analogous way to that done
in section III \cite{billyard}.
The equilibrium set $L^+$: 
$x^2 + y^2 = 1 , z=u=v=0$ still exists, and since the eigenvalues 
associated with 
($u$ and) $v$ are (both) negative, part of $L^+$ 
will act as sinks and the non-hyperbolic
point $x = -1 /\sqrt{3} , y = -\sqrt{2/3},  z=u=v=0$ is clearly a saddle.
These are local results and are valid in all cases.

There also exists an attracting equilibrium point $W$:
\begin{equation}
  x = -\sqrt{3}\left( 12 + c^2 \right)D^2\ , \ y = 8c\sqrt{6}D^2\ ,
  \ z = -192CD^4\ , \ u = v = 0\ , 
\end{equation}
with eigenvalues
$$
-2CD^2\ , \ -(C \pm \sqrt{C(17c^2+924)})D^2\ ,
-4(12+c^2)D^2\ , \ -6(12+c^2)D^2\ ,
$$
where $C \equiv (c+6)(c-6)$ and $D^{-2} \equiv 108+c^2$.
Note that this point corresponds to an exact self-similar collapsing cosmological solution
with non--trivial modulus and four--form fields.  The value of $\Omega$ as a function of $c$ is given by
\begin{equation}
  \Omega = \frac{-2C(60+c^2)}{(108+c^2)^2} ,
\end{equation}
which implies that $-6\leq c \leq
6$ in order for $\Omega\geq0$. When $c=\pm6$, the point $W$ is a part of the 
equilibrium set $L^+$; 
in fact, it becomes just the non-hyperbolic point $P$ discussed previously.  
Note that it is also a part of the invariant submanifold
$cx - 6\sqrt{2}y + \sqrt{3}c = 0; u=v=0$, which generalizes 
the invariant submanifold $K$ defined in Eq. (\ref{eq:nontriv}).

There are two non-flat ($u\neq0$) vacuum equilibrium points with a vanishing
cosmological constant $v=0$,
one of which is a source and the other a saddle.
There is also a non-hyperbolic vacuum equilibrium point with a non-vanishing
cosmological constant ($v=1$) with
$x=y=z=u=0$, which appears to be a source.
In addition, we can find monotone functions in the boundary submanifolds;
indeed, the boundary 
$\Omega = u=v=0$ and the boundary submanifold
$z = u=v=0$ can be solved
exactly in terms of the variables of the state space. Exact solutions
of the equations of motion for particular values of $c$ 
can also be found.

However, the primary motivation for these comments is to 
emphasize two important points regarding 
the very interesting dynamics of the M--theory cosmologies
studied earlier. First, we note that the conclusions obtained 
for the spatially curved models are 
robust when additional physical fields (e.g., a $\Lambda$ term) 
are included. Second, and perhaps
more importantly,
we see that the value $c=-6$ is a {\em bifurcation value} in the analysis
of general models with arbitrary coupling, $c$. In this context, 
therefore, the M-theory 
cosmological models we have studied exhibit rather 
unique dynamics. 

\section{Discussion}

In this paper we have presented a complete dynamical analysis of spatially 
flat, four--dimensional 
cosmological models derived from the M-theory and type IIA string 
effective actions. 
We have shown that models generically spiral away from a source $P$, 
undergoing bounces due to the
interplay between the NS-NS two-form potential 
and the RR three-form potential.
Eventually, they evolve 
towards dilaton-moduli-vacuum solutions 
with trivial form fields (corresponding to the
sinks on $L^+$). We note the important dynamical 
result that $\varphi'$ is monotonic.

Thus, the 
form fields that arise as massless excitations in the type IIA
superstring spectrum, or equivalently from the three--form potential 
of eleven--dimensional supergravity, may have important consequences 
in determining initial and final conditions in string and M--theory
cosmologies, even though they are dynamically negligible 
in the early-- and late--time limits. In particular, the point 
$P$ is the only source in the system. It can be interpreted in the string 
context as the isotropic, ten-dimensional solution. Alternatively, 
it represents the Taub form of flat space when viewed in terms 
of eleven--dimensional variables. 

When the effects of spatial curvature are included, we 
obtained the local result that the point $P$ becomes a saddle. On 
the other hand, the dilaton-moduli-vacuum 
solutions with trivial form fields
are generic attracting solutions. 
In the analysis of the negatively-curved models,  we
found that the early time attractor (the source $R$)
has non-zero curvature, implying that 
spatial curvature is dynamically important at early times in these 
examples.

This work can be generalized in a number of ways.
We considered a specific compactification from eleven to four dimensions, 
where the topology of the internal dimensions was assumed to be 
a product space consisting of a circle and an isotropic six--torus. 
We emphasize, however, 
that the analysis also applies to compactifications on 
a Calabi--Yau three--fold since the gauge fields arising from the 
higher--dimensional metric have been ignored \cite{kko}. 
The qualitative analysis may be readily extended to 
compactifications on a general, rectilinear torus $S^1 \times 
\ldots \times S^1$. After suitable redefinitions of the additional 
moduli fields that subsequently arise, 
the dimensionally reduced action can be expressed precisely in the form of 
Eq. (\ref{solitonaction}), with the inclusion of a set of massless 
scalar fields in the NS--NS sector. In particular, 
the compactification  on $T^4\times T^2  \times S^1$, where 
$T^n$ represents the isotropic $n$--torus, is 
relevant to compactifications involving the four--dimensional space 
$K3$ \cite{K3}. This space has played an important r\^{o}le in establishing 
various string dualities \cite{Sen98}. 
It is the simplest four--dimensional, Ricci--flat 
manifold after the torus \cite{aspinwall} and 
may be approximated by the orbifold $K3 \approx T^4/
{\rm Z}_2$ \cite{K31}. 

Moreover, the effects of spatial anisotropy in the 
spacetime $M_4$ 
can also be considered by introducing two, uncoupled moduli fields 
into  the NS--NS sector of the reduced action 
(\ref{1daction}). In this context, such fields 
parametrize the shear in the cosmologies. 
When these fields are non--trivial, the models represent the
class of isotropic curvature cosmologies and 
correspond to Bianchi type I, V and IX universes \cite{BCL,iscu}. 
It would be interesting to consider these generalizations further.

\acknowledgments

APB is supported by Dalhousie University, AAC is supported by the
Natural Sciences and Engineering Research Council of Canada (NSERC),
JEL is supported by the Royal Society and USN is supported by
G\r{a}l\"{o}stiftelsen, Svenska Institutet, Stiftelsen Blanceflor and
the University of Waterloo USN. We thank N. Kaloper and I. Kogan 
for helpful communications.


\references

\bibitem{gsw} M. B. Green, J. H. Schwarz, and E. Witten,
{\em Superstring Theory}, in 2 vols., (Cambridge University 
Press, Cambridge, 1987); J. Polchinski, {\em String Theory}, in 2 vols., 
(Cambridge University Press, Cambridge, 1998). 

\bibitem{Witten95} E. Witten, Nucl. Phys. {\bf B443}, 85 (1995). 

\bibitem{Sen98} A. Sen, hep-th/9802051.

\bibitem{Townsend95} P. Townsend, Phys. Lett. {\bf B350}, 184 (1995). 

\bibitem{hw} A. Lukas, B. A. Ovrut, and D. Waldram, hep-th/9806022; 
A. Lukas, B. A. Ovrut, and D. Waldram, hep-th/9812052; H. S. Reall, 
Phys. Rev. {\bf D59}, 103506 (1999); K. Benakli, Int. J. Mod. Phys. 
{\bf D8}, 153 (1999); K. Benakli, Phys. Lett. {\bf B447}, 51 (1999). 

\bibitem{others} H. Lu, S. Mukherji, and C.N. Pope, hep-th/9612224; 
A. Lukas, B. A. Ovrut, and D. Waldram, Nucl. Phys. {\bf B495}, 365 (1997); 
H. Lu, S. Mukherji, 
C. N. Pope, and K. -W. Xu, Phys. Rev. {\bf D55}, 7926 (1997); 
A. Lukas, B. A. Ovrut, Phys. Lett. {\bf B437}, 291 (1998); 
H. Lu, J. Maharana, S. 
Mukherji, and C. N. Pope, Phys. Rev. {\bf D57}, 2219 (1998); 
A. Lukas, B. A. Ovrut, and D. Waldram, Nucl. Phys. {\bf B509}, 169 (1998); 
M. Bremer, M. J. Duff, 
H. Lu, C. N. Pope, and K. S. Stelle, Nucl. Phys. {\bf B543}, 321 (1999); 
M. J. Duff, P. Hoxha, H. Lu, R. R. Martinez--Acosta, and C. N. Pope, 
Phys. Lett. {\bf B451}, 38 (1999); S. W. Hawking and H. S. Reall, Phys. 
Rev. {\bf D59}, 023502 (1999). 

\bibitem{kko} N. Kaloper, I. I. Kogan, and K. A. Olive, Phys. 
Rev. {\bf D57}, 7340 (1998); Erratum, ibid. {\bf D60}, 049901 (1999). 

\bibitem{fvm} A. Feinstein and M. A. Vazquez--Mozo, hep-th/9906006.

\bibitem{CJ79} E. Cremmer, B. Julia, and J. Scherk, Phys. Lett. {\bf B76}, 
409 (1978); E. Cremmer and B. Julia, Phys. Lett. {\bf 
B80}, 48 (1978); E. Cremmer and B. Julia,  Nucl. Phys. {\bf B156}, 141 
(1979). 

\bibitem{citation} L. Castellani, P. Fre, F. Giani, K. Pilch, and P. 
van Nieuwenhuizen, Ann. Phys. {\bf 146}, 35 (1983). 

\bibitem{effectivecircle} I. C. Campbell and P. C. West, Nucl. Phys. 
{\bf B243}, 112 (1984); F. Giani and M. Pernici, Phys. Rev. 
{\bf D30}, 325 (1984); M. Huq and M. A. Namazie, Class. Quantum 
Grav. {\bf 2}, 293 (1985); Erratum, ibid. {\bf 2}, 597 (1985). 

\bibitem{stw91} A. Shapere, S. Trivedi, and F. Wilczek, Mod. 
Phys. Lett. {\bf A6}, 2677 (1991); A. Sen, Mod. Phys. Lett. {\bf A8}, 
2023 (1993). 

\bibitem{Qansatz} P. G. O. Freund and M. A. Rubin, Phys. Lett. {\bf B97}, 
233 (1980); F. Englert, Phys. Lett. {\bf B119}, 339 (1982). 

\bibitem{tsy} A. A. Tseytlin, Int. J. Mod. Phys. {\bf D1}, 223 (1991). 

\bibitem{Veneziano91} G. Veneziano, Phys. Lett. {\bf B265}, 287 (1991).

\bibitem{WEBOOK} J. Wainwright and G. F. R. Ellis, {\em Dynamical Systems 
in Cosmology} (Cambridge University Press, Cambridge, 1997). 

\bibitem{coplahwan} E. J. Copeland, A. Lahiri, and D. Wands, 
Phys. Rev. {\bf D50}, 4868 (1994). 

\bibitem{kaloper} N. Kaloper, Phys. Rev. {\bf D55}, 3394 (1997). 

\bibitem{cowlupop} H. Lu, C. N. Pope, E. Sezgin, and K. S. Stelle, 
Nucl. Phys. {\bf B456}, 669 (1995); 
H. Lu, C. N. Pope, E. Sezgin, and K. S. Stelle, 
Phys. Lett. {\bf B371}, 46 (1996); M. Cvetic and H. H. Soleng, 
hep-th/9604090;
P. M. Cowdall, H. Lu, C. N. Pope, K. S. Stelle, and 
P. K. Townsend, Nucl. Phys. {\bf B486}, 49 (1997). 

\bibitem{pbb} M. Gasperini and G. Veneziano, Astropart. Phys. {\bf 1}, 
317 (1993).

\bibitem{BCL} A. P. Billyard, A. A. Coley, and J. E. Lidsey, gr-qc/9907043;
A.P. Billyard, A. A. Coley and J. E. Lidsey, ``Qualitative Analysis 
of Isotropic Curvature String Cosmologies'' (1999). 

\bibitem{billyard} A. P. Billyard, Ph. D Thesis (1999).

\bibitem{K3} M. J. Duff, B. E. W. Nilsson, and C. N. Pope, 
Phys. Lett. {\bf B129}, 39 (1983); H. Lu, C. N. Pope, and K. S. Stelle, 
Nucl. Phys. {\bf B548}, 87 (1999). 

\bibitem{aspinwall} P. S. Aspinwall, hep-th/9611137. 

\bibitem{K31} D. N. Page, Phys. Lett. {\bf B80}, 
55 (1978); G. W. Gibbons  and C. N. Pope, Commun. Math. Phys. 
{\bf 66}, 267 (1979).

\bibitem{iscu} M. A. H. MacCallum, {\em Cargese Lectures in Physics} ed. E. 
Schatzman (Gordon and Breach, New York, 1973).
\end{document}